# Flexoelectronic doping of the degenerate silicon and the correlated electron behavior

Paul C. Lou[1, ‡], Anand Katailiha[1, ‡], Ravindra G. Bhardwaj[1, ‡], Ward Beyermann[2], Dheeraj Mohata[3], and Sandeep Kumar[1,*]

[1] Department of Mechanical Engineering, University of California, Riverside, CA 92521, USA

[2] Department of Physics and Astronomy, University of California, Riverside, CA 92521, USA

[3] Global Communication Semiconductors LLC, Torrance, CA 90505

[*] Corresponding author

[‡] Equal contribution

Email: skumar@engr.ucr.edu

**Abstract**

In metal/degenerately doped silicon bilayer structure, the interfacial flexoelectric effect due to strain gradient leads to charge carrier transfer from metal layer to the silicon layer. This excess charge carrier concentration is called flexoelectronic doping or flexoelectronic charge transfer, which gives rise to an electronically polarized (order of magnitude larger than ferroelectric materials) silicon layer. In the transport measurements, the charge carrier concentration in silicon is found to increase by two orders of magnitude due to flexoelectronic doping, which changes the Fermi level and the Hall response. The flexoelectronic charge accumulation modifies the electron-electron and the electron phonon coupling, which gives rise to Mott metal-insulator transition and magnetism of phonons, respectively. The coexistence of flexoelectronic polarization and magnetism gives rise to a new class of materials called electronic multiferroics. By controlling the flexoelectronic doping, material behavior can potentially be engineered for quantum, spintronics and electronics applications in semiconductor materials.

# I. Introduction

Si is pre-eminent semiconductor that has led to the modern electronics revolution. The ability to dope Si with impurity atoms has allowed control of its conductivity from intrinsic to metal like enabling a multitude of device and circuit applications. Recently, Wang et al. [1] demonstrated an alternative method to control interfacial electronic transport using large inner crystal (or bulk) flexoelectric charge separation. The flexoelectric charge separation can also, potentially, be relevant to spintronics, spin caloritronics and other future applications, and not just only electronics. In addition, due to strain, the interfacial flexoelectric polarization[2-4] in thin film heterostructures is expected to be much larger than the bulk counterpart and that too can contribute towards the bulk material response. Hence, the strain gradient can potentially be one of the tools to tailor the properties of a heterostructure for the desired material response.

In a recent study, a large spin-Hall effect was reported in degenerately doped Si under an applied strain gradient, which was an order of magnitude larger than that observed in Pt and similar to the topological insulators surface states[5]. In another study, a large magnetic moment of 1.2 $\mu_B$/atom due to phonons was reported in the flexoelectronic Si[6]. The measured phonon magnetic moment in highly doped Si was four orders of magnitude larger than that predicted according to the circular motion of ions[7-9]. These responses were attributed to the flexoelectronic polarization in the Si thin films, but the mechanistic origin of it was not explained. Also, according to the current understanding, no polarization should arise in highly doped semiconductor samples since mobile charge carriers are expected to quench them, which is also supported by the study of Wang et al.[1] This contradiction in the observed behavior was attributed to the lack of understanding of the material physics of the underlying bulk and interfacial flexoelectronic behavior in doped semiconductors. This motivated our work since it could give rise to polarization in a conducting material, which could open further opportunities for fundamental and applied materials research.

II. Hypothesis

In semiconductors, strain gradient results in slope in the valence band maxima, the conduction band minima, and the charge carrier mobilities in addition to the band gap. In the case of the lightly doped Si, the slope of the band structure typically results in charge separation and bulk flexoelectric effect as shown in Figure 1 (a)[10]. However, in the case of highly doped Si, the free charge carriers are expected to screen the flexoelectric effect as shown in Figure 1 (a) and thus no net polarization can be detected. In contrast, an interfacial flexoelectric effect[1] appears at the interface in the metal-Si (highly doped) bilayer structure as shown in Figure 1 (b). The interfacial flexoelectric effect[3,11] results in free charge carrier injection from metal layer to the impurity band of the degenerate Si layer as shown in Figure 1 (c). This charge carrier injection is expected to give rise to a gradient of the charge carrier concentration as shown in Figure 1 (d). As a consequence, the flexoelectronic polarization, equal and opposite to the interfacial flexoelectric polarization, is expected to arise in the bulk of the Si thin film as shown in Figure 1 (c,d). This effect can be considered an inverse of the behavior reported by Wang et al.[1] since the interfacial effect drives the bulk response. In thick Si samples, the flexoelectronic charge carrier gradient is expected to be nearer to the interface only as shown in Figure 1 (d). In case of lightly doped thick Si samples, the behavior will converge with the response reported by Wang et al.[1], since there is no impurity band for flexoelectronic doping. Whereas in the thin film (micro/nano) Si samples, the average charge carrier density in the highly doped Si layer will increase, for example, from say $5\times10^{19}$ cm$^{-3}$ to $7\times10^{19}$ cm$^{-3}$ in addition to the charge carrier concentration gradient as shown in Figure 1 (d). We call this phenomenon as flexoelectronic doping of the Si thin films or flexoelectronic charge transfer that is driven by the free electronic charge carriers transferred from metal to Si.

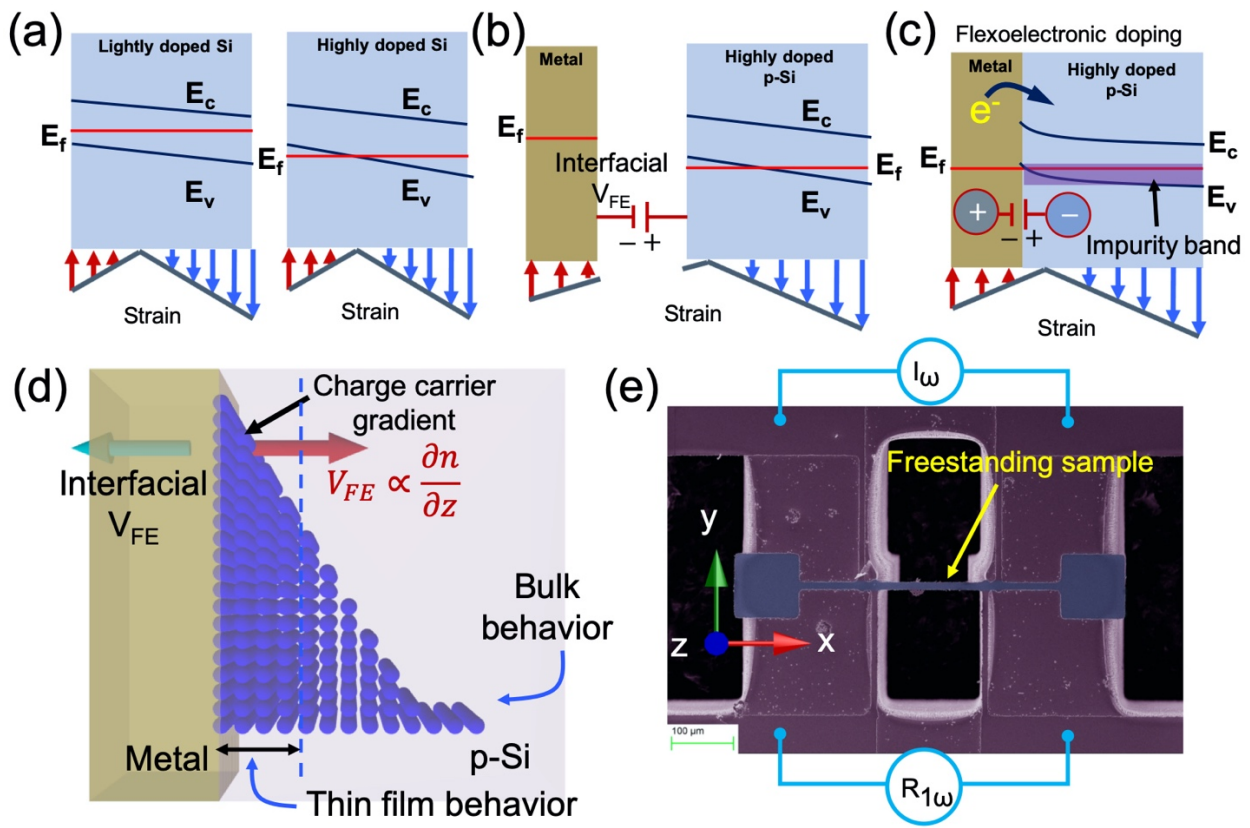

Figure 1. (a) schematic showing approximate band diagram of p-doped Si thin film from flexoelectric charge separation due to linearly varying strain (constant strain gradient). (b) Schematic showing the band structure of metal/p-Si bilayer structure with interfacial flexoelectric voltage due to linearly varying strain (constant strain gradient), (c) schematic showing the charge carrier injection and flexoelectronic doping in the Py/Si bilayer (cross-

section) thin film. (d) A schematic showing the charge carrier concentration gradient in the cross-section of the p-Si layer with flexoelectronic polarization equal and opposite to the interfacial flexoelectric polarization. (e) A false color scanning electron micrograph showing the experimental scheme and top view of the freestanding sample structure.

Since Si is no longer charge neutral, flexoelectronic doping is expected to lead to renormalization of the electronic wave function as well as phonons, which would modify the electron-electron and electron-phonon coupling. Hence, both weakly and strongly correlated electronic behavior could potentially be induced in Si thin films. In this study, we report first experimental evidence of the large flexoelectronic doping of the degenerate Si layer in the metal-Si heterostructures. This flexoelectronic doping results in about two orders increase in charge carrier concentration from $8.73 \times 10^{18}$ cm$^{-3}$ to $1.55 \times 10^{21}$ cm$^{-3}$. This charge accumulation further gives rise to strong electron-electron Coulomb repulsion, which results in Mott metal insulator transition (MIT) in the Si layer. Additionally, the superposition of the flexoelectronic polarization and phonons gives rise to magnetic moment in the Si layer and giant magnetoresistance (GMR) response in the sample. As a consequence, metal/Si heterostructures can be considered electronic multiferroic materials.

### III.     Experimental results and discussion

To study the flexoelectronic doping in Si thin films, we chemically etched a 2 µm single crystal p-Si (Boron doped) device layer (0.001-0.005 Ω cm) of the SOI wafer to achieve the thickness closer to ~400 nm by successively oxidizing (SiO$_2$ formation using thermal oxidation) and, then, etching the thermal oxide using hydrofluoric (HF) acid[12]. This method is designed such that larger strain gradient effects are achieved in Si. The device structure and the backside Si underneath the sample is etched using Deep reactive ion etching (DRIE). Then, the MgO (1.8 nm) and $Ni_{80}Fe_{20}$ (Py) (25 nm) layers were deposited using sputtering and e-beam evaporation, respectively, on the etched freestanding p-Si thin films structure as shown in Figure 1 (e). The metal and oxide depositions are expected to induce large strain gradient in the Si layer as well as at the interface due to thermal mismatch stresses. Using a qualitatively similar technique, photonics in inhomogeneously strained Si have already been demonstrated[13,14]. Previously, a strain of 4% was estimated using high resolution transmission electron microscope (HRTEM) in 2 µm p-Si near the MgO interface[5]. However, HRTEM sample fabrication can also induce additional strain. Hence, the strain distribution and strain gradient were not measured in this study. The MgO layer is used to prevent any metal diffusion into Si. Based on previous results, the MgO layer was not necessary and was introduced to remove the potential alloying effects at the interface[15,16]. However, the flexoelectric polarization of MgO layer also contribute towards flexoelectronic doping.

### A. Flexoelectronic doping in Py/MgO/p-Si (400 nm) sample

In the first study, we measured the longitudinal resistance as a function of temperature from 350 K to 5 K and current bias from 10 µA to 2 mA as shown in Figure 2 (a) inside a Quantum Design PPMS chamber using standard lock-in technique. The sample (sample 1) dimensions are: length- 160 µm, width- 11.7 µm and thickness- 400 nm. These responses are significantly different from the Py and p-Si responses, individually[12]. The resistance of the sample reduces as the current bias is increased to 2 mA as shown in Figure 2 (a). The mechanism of the reduction is attributed to the increased strain gradient due to thermal mismatch stresses from self-heating of the freestanding sample[5]. It is noted that Joule heating due to larger current will lead to increase in resistance as opposed to the observed response, which eliminated Joule heating as the probable cause of the observed behavior. The resistance responses exhibit an MIT below 100 K as shown in Figure 2 (a,b). The residual resistance after MIT is larger at higher currents as shown in Figure 2 (b). The MIT response cannot arise in the Py layer having metallic bonding. Hence, the Si layer is expected to become an insulator at reduced temperatures. As a consequence, the residual resistance at 5 K is expected to be from the Py layer only, as shown in Figure 2 (b).

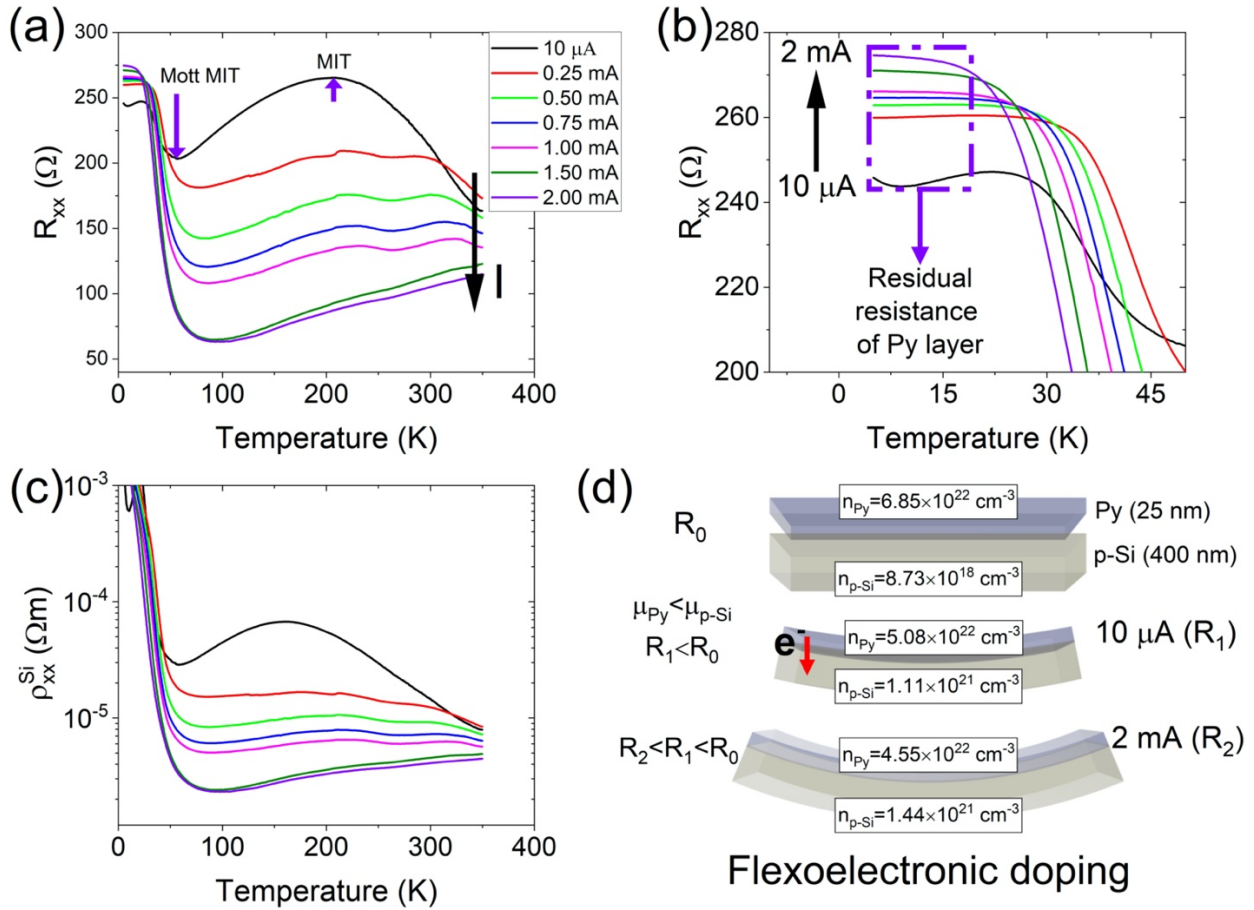

Figure 2. (a) the longitudinal resistance response in Py (25 nm)/MgO/p-Si (400 nm) composite sample as a function of temperature and current bias from 10 µA to 2 mA, (b)

the resistance response between 50 K and 5 K showing the Mott metal insulator transition in the Si layer and residual resistance of Py layer, (c) the estimated resistivity of the Si layer in the composite sample after subtracting Py response, and (d) a schematic showing the change in the average charge carrier concentration from flexoelectronic doping of the Si layer due to charge carrier transfer from Py layer to Si layer.

The Py resistance response as a function of temperature from 300 K to 5 K has been reported previously[12]. From previous data, the resistance of the Py layer in the current sample is expected to be ~182.6 Ω at 5 K as shown in Supplementary Figure S1, whereas the residual resistance is ~ 245.8 Ω at 10 µA. The ratio of the measured and expected resistance at 5 K is calculated and used as a multiplication factor to estimate the Py resistance behavior for measurement at each current. We, then, subtracted the Py resistance response from the sample resistance and extracted the resistivity behavior of Si layer, which is shown in Figure 2 (c). The estimated resistivity of the p-Si layer is an order of magnitude larger than the wafer device layer resistivity, which is a clear indicator of flexoelectronic doping. The resistance in Py/MgO/p-Si (400 nm) sample is even smaller than that in Py/MgO/p-Si (2 µm) sample[17].

The flexoelectronic doping is attributed to charge carrier transfer from Py to p-Si layer. We divided $n_{Py}= 6.85\times10^{22}$ cm$^{-3}$ with the multiplication factor (calculated earlier) to estimate the new charge carrier concentration as well as the change in the charge carrier concentration for each current dependent measurement presented in Figure 2 (a). For example- the difference in charge carrier concentration is expected to be $1.96\times10^{22}$ cm$^{-3}$ at 10 µA of current in Py layer. The difference in the charge carrier concentration is then transferred to the Si layer. As a result, the new charge carrier concentration of Si at 10 µA of current increases to $1.11 \times 10^{21}$ cm$^{-3}$ as shown in Figure 2 (d) as compared to ~$8.73 \times 10^{18}$ cm$^{-3}$ measured from p-Si (400 nm) sample as shown in Supplementary Figure S2. The flexoelectronic doping value is similar to one of the highest doping levels achieved in Si[18] using impurities. The increased sample bending from larger current bias results in larger flexoelectronic doping as shown in Figure 2 (d) and Supplementary materials. The mobility of the charge carrier in p-Si (~131.4 cm$^2$/(V.s)) is two orders of magnitude larger than that in Py (~1.7 cm$^2$/(V.s)) as shown in Supplementary Figure S2, which is the underlying cause of the reduction in the overall resistance due to flexoelectronic doping. The charge carrier concentration estimated in this study is an average value and a gradient of charge carrier is expected to exist due to gradient in the band structure as stated earlier. It is noted that strain gradient accompanied gradient of mobilities through the thickness of the sample. However, its effect on charge transport and flexoelectronic doping is not characterized.

The change in charge carrier concentration is calculated for each current as shown in Supplementary Table S1. Based on the number of charge carrier injected, we estimated the maximum charge accumulation in Si layer to be 71.1 C/m$^2$ at 10 µA of current, which is three orders of magnitude larger than the value reported recently by Wang et al.[1] and is expected to increase to 79.4 C/m$^2$ at 2 mA due to larger strain gradient. This value is also an order of magnitude larger than ferroelectric polarization (1.5 C/m$^2$)[19] even though a direct comparison is not possible due to electronic nature of the charge accumulation in this study. This large charge accumulation is believed to be the underlying cause of the phonon magnetic moment of 1.2 µ$_B$/atom in flexoelectronically doped Si[6] as compared to $10^{-4}$ $\mu_B$[7] predicted by the dynamical multiferroicity. The charge accumulation also shows that the flexoelectric effect at interfaces can be very large and should not be ignored.

Our resistance measurements showed an MIT behavior below 100 K, as stated earlier. Another MIT like behavior is observed at 200 K, but only in the measurement at 10 µA current bias. In B-doped Si, the MIT occurred as a function of the doping at ~3-4×$10^{18}$ cm$^{-3}$.[20,21] However, in the present study, the p-Si layer is already degenerately doped (~8.73×$10^{18}$ cm$^{-3}$) as shown in Supplementary Figure S2. Hence, flexoelectronic charge accumulation is expected to be the underlying cause of MIT behavior. The excess charge carriers from flexoelectronic doping are expected to modify the electron-electron Coulomb repulsion and possibly open a gap, which led to Mott transition[20,22] below 100 K as shown in Figure 2 (a-c). At 10 µA current, weaker Coulomb repulsion drove a continuous MIT at 200 K, as shown in Figure 2 (a), but it is suppressed due to self-heating of the sample at larger current bias. The MIT response is not observed in Py/MgO/p-Si (2 µm) samples due to smaller strain gradient and consequently smaller flexoelectronic doping[17]. At this time, the band structure of the flexoelectronically doped Si is not available, which is part of the future theoretical work. The observed Mott MIT is an example of correlated electronic behavior due to change in electron-electron interactions from flexoelectronic doping. It is noted that the flexoelectronic doping level is similar to the doping level achieved in superconducting Si[18]. However, the charge polarization from flexoelectronic doping results in Mott MIT at much higher temperature instead of superconductivity at $T_c$=0.35 K[18]. While we did not study superconductivity in our samples, but further research may make it feasible.

## B. Flexoelectronic spin current in Py/MgO/p-Si (400 nm) sample

In the absence of an external magnetic field, the charge carrier injection from the ferromagnetic (Py) layer to p-Si layer is not spin polarized and only an equilibrium charge carrier injection ($j_c$) is expected to exist. However, an applied magnetic field results in spin polarization of the charge carriers in the ferromagnetic layer. As a consequence, charge carrier injection will be accompanied by a flexoelectronic spin current ($j_s$) as shown in Figure 3 (a). To study it, we measured the longitudinal resistance as a function of temperature for an applied out-of-plane magnetic field of 0 T and 1 T at a current bias of 1 mA in the sample 1 as shown in Figure 3 (a). The longitudinal resistance is ~54.5% smaller at ~85 K in the case of applied magnetic field as compared to zero field as shown in Figure 3 (a). This reduction in resistance is attributed to reduced scattering due to larger ferromagnetic exchange interactions and reduced Coulomb repulsion from enhanced spin polarization in the Si layer. The reduction in resistance can also be attributed to spin-to-charge conversion from flexoelectronic spin current. The Mott-MIT response as well as residual resistance are similar. The large flexoelectronic spin current is also potentially the reason for the large spin-Hall effect as well as the spin-Seebeck effect reported in Si thin films[5,16,23].

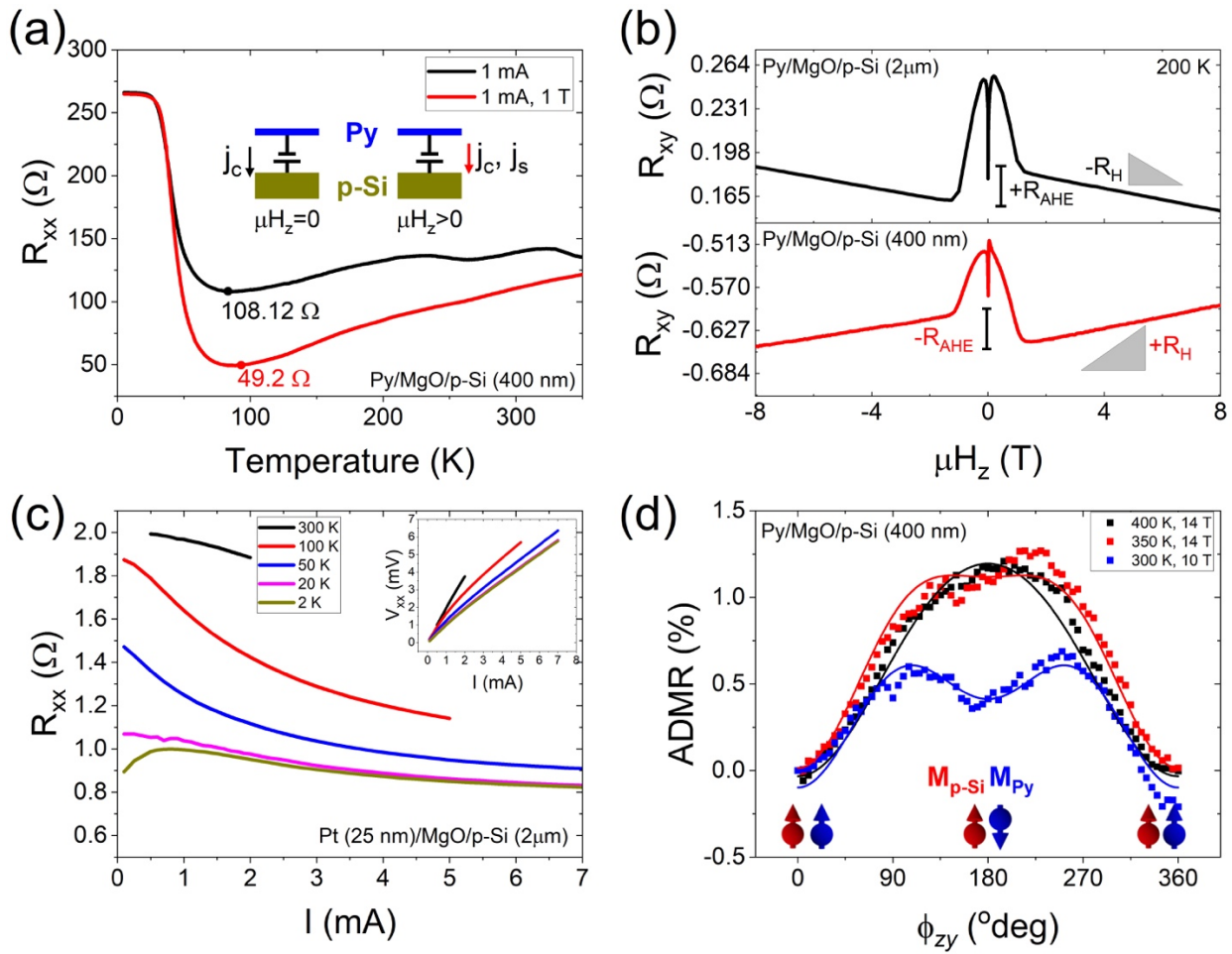

Figure 3. (a) The longitudinal resistance response in Py (25 nm)/MgO/p-Si (400 nm) composite sample at an applied current bias of 1 mA for a zero applied magnetic field and 1 T out of plane magnetic field. (b) The Hall resistance measurement for an applied magnetic field of 8 T to -8T at 200 K showing negative Hall resistance in Py (25 nm)/MgO (1.8 nm)/p-Si (2 μm) in top panel and positive Hall resistance in Py (25 nm)/MgO(1.8 nm)/p-Si(400 nm) in bottom panel. (c) The resistance as a function of current measured at 300 K, 100 K, 50 K, 20 K and 2 K in a Pt (25 nm)/MgO (1.8 nm)/p-Si (2 μm) sample. Inset shows voltage vs the current response. (d) The angle dependent magnetoresistance measured in freestanding Py/MgO/p-Si (400 nm) sample showing the GMR response, which decreases with temperature. The solid line shows the curve fit.

### C. Hall effect measurement in Py/MgO/p-Si samples

The transfer of the large number of charge carriers from Py layer to p-Si layer is expected to change the Fermi level and, as a consequence, Hall resistance behavior. We, then, measured the Hall resistance as a function of magnetic field from 8 T to -8 T at 200 K in Py(25 nm)/MgO(1.8 nm)/p-Si (400 nm) (sample 2) and compared it with a Py(25 nm)/MgO(1.8 nm)/p-Si (2 µm) sample (sample 3). The sample dimensions are: length- 40 µm and width-14 µm. In the 2 µm p-Si sample, the Hall resistance is negative corresponding to electrons and the anomalous Hall resistance is positive as shown in top panel in Figure 3 (b). Whereas in the 400 nm p-Si sample, the sign of both the Hall resistance and the anomalous Hall resistance are reversed as shown in bottom panel in Figure 3 (b). The positive Hall resistance corresponded to holes, which is expected to arise due to flexoelectronic charge carrier transfer from Py to p-Si. Similarly, the anomalous Hall response in Py is due to an intrinsic mechanism[24]. As a consequence, its sign reversal also corresponds to the change in the type of the charge carrier from electrons to holes. A similar electron to hole transition was previously reported in Pt/MgO/p-Si sample[6]. Since the MgO and Py layer thicknesses are same in both the samples, the strain gradient is expected to be smaller in the 2 µm p-Si sample as compared to the 400 nm p-Si sample because the critical stress required for beam buckling will be larger in the thicker sample. As a consequence, the smaller flexoelectronic charge carrier transfer will not change the sign of Hall resistance in case of 2 µm p-Si in sample 3 as opposed to the larger strain gradient and flexoelectronic doping in case of 400 nm p-Si layer in sample 2. This measurement provides a second proof of flexoelectronic doping. In addition, it shows that the electronic properties can potentially be controlled using strain gradient even in the case of metal and degenerately doped Si heterostructures.

### D. Flexoelectronic doping in Pt/MgO/p-Si (2 µm) sample

For additional experimental proof of the flexoelectronic doping, we, then, fabricated a Pt (25 nm)/MgO (1.8 nm)/ p-Si (2µm) thin film Hall bar sample (sample 4), which allowed us to uncover the effect of electronic source metal layers. The sample dimensions are: length- 40 µm and width- 14 µm. We measured the resistance as a function of applied longitudinal current and as a function of temperature as shown in Figure 3 (c). The resistivity of the Pt thin film is expected to be $2.57\times10^{-7}$ $\Omega$-m [6] whereas the resistivity of the Si device layer is expected to be $1-5\times10^{-5}$ $\Omega$-m. However, the resistance of the sample at 300 K is measured to be 2.02 $\Omega$ as compared to 9.61 $\Omega$ (using the smallest resistivity of the p-Si layer). It is noted that even if we used the bulk resistivity value for Pt ($1\times10^{-7}$ $\Omega$ m), the measured resistance could not be lower than 6.35 $\Omega$. The new resistivity of the Si layer needed to simulate the observed sample resistance is estimated to be $1.518\times10^{-6}$ $\Omega$ m, which could only occur due to charge carrier transfer from Pt layer to the Si layer due to strain gradient. This value of the resistivity is smaller than the estimated value in p-Si (400 nm) sample as shown in Figure 2 (c). The melting point of Pt is significantly higher (than Py) and, as a consequence, the residual stresses in the Pt layer are expected to be larger, which led to larger flexoelectronic charge transfer and doping. This reduction in resistance occurs since the mobility of charge carriers in p-Si is an order of magnitude larger than that of metals as stated earlier. The larger current should increase the resistance of the sample due to self-heating; instead resistance of the sample decreases as shown in Figure 3 (c). This behavior is similar to the previously reported spin-Hall magnetoresistance response as a function of current[5]. The voltage-drop as a function of current also shows a non-linear behavior as shown in the inset of Figure 3 (c). This behavior is again attributed to the larger strain gradient from self-heating and larger flexoelectronic doping of Si layer. This is the third experimental evidence of flexoelectronic doping of Si layer in a metal/Si bilayer structure. In the Py/p-Si (400 nm) sample, the charge carrier injection also led to exchange interactions, which led to Mott MIT. However, we do not observe Mott MIT in the Pt/p-Si (2 μm) sample even though the flexoelectronic doping levels are expected to be larger. We attribute this behavior to the absence of ferromagnetic proximity-induced exchange interactions. This measurement showed that by controlling the kind of flexoelectronic doping; the correlated electron behavior can potentially be controlled. We also propose that flexoelectronic doping from different metals (lighter or heavier) will lead to different correlated electron behavior, which is part of the future work.

### E. Giant magnetoresistance response in Py/MgO/p-Si (400 nm) sample

The superposition of flexoelectronic polarization and temporal polarization of phonons gives rise to temporal magnetic moment due to dynamical multiferroicity, as reported previously. The dynamical multiferroicity in Si thin films under an applied strain gradient[6] can be described as:

$$\boldsymbol{M}_t \propto \boldsymbol{P}_{FE} \times \partial_t \boldsymbol{P} \qquad (1)$$

where $\boldsymbol{M}_t$, $\partial_t \boldsymbol{P}$ and $\boldsymbol{P}_{FE}$ $\left(\propto \frac{\partial n}{\partial z}\right)$ are temporal magnetic moment, time-dependent polarization of optical phonons and flexoelectronic effect[1] from charge carrier concentration gradient, respectively. To measure the magnetism, we measured the angle-dependent magnetoresistance of the sample as a function of temperature in the sample 1 as shown in Figure 3 (d). The angle-dependent magnetoresistance in zy-plane (plane perpendicular to current direction) at 400 K and 14 T showed a $\cos\theta_{zy}$ angular symmetry, which is attributed to the GMR for a current in-plane (CIP) geometry. The GMR is ~1.22% at 14 T magnetic field as shown in Figure 3 (d). Based on symmetry, the magnetic moment in the Si layer is expected to be in the +z-direction and remained fixed, while the Py magnetic moment always aligned with the magnetic field giving rise to the GMR response as shown in Figure 3 (d). The response at 350 K is similar, but the magnitude of the GMR reduces to 1.11% at 14 T magnetic field as shown in Figure 3 (d). At 300 K, the GMR response decreases to ~0.51% at 10 T magnetic field. The additional $\sin^2\theta_{zy}$ response at 350 K and 300 K is attributed to anisotropic magnetoresistance (AMR) from the Py layer. The reduction in the GMR response as a function of temperature is attributed to the magnetism of phonons or dynamical multiferroicity. The reduction in temperature causes the reduction in polarization of the optical phonons, which in turn reduces the temporal magnetic moment. The observed magnetic behavior is a second example of correlated electron behavior due to modification in electron-phonon coupling[25] from flexoelectronic doping. In addition, the coexistence of flexoelectronic polarization and magnetism is attributed to dynamical multiferroicity. As a consequence, the metal/Si (doped) heterostructures can be called as electronic multiferroic, which is a new class of materials.

While we have demonstrated flexoelectronic doping and resulting magnetism and Mott MIT but the quantitative relationship with the strain gradient is not measured. Currently, there is no known method to address it since the samples are conducting.

Additionally, the flexoelectronic charge transfer create a charge deficient metal layer, which could give rise to correlated electron behavior in metal layers as well. The flexoelectronic behavior in metal layer can have potential application in catalysis. The future theoretical studies are also needed to uncover the changes in the band structure, renormalization of the electronic wave function and phonons, which is a modelling challenge due to lack of charge neutrality in the flexoelectronic doping.

## IV. Conclusion

In conclusion, we present the first experimental evidence of large flexoelectronic doping in a degenerately B-doped Si. The flexoelectronic doping arises from the charge carrier injection from the metal layer to the Si layer due to interfacial flexoelectric effect. The flexoelectronic doping gives rise to electron-electron Coulomb repulsion and Mott MIT. Additionally, flexoelectronic doping gives rise to magnetic moment in Si due to electronic dynamical multiferroicity. The discovery of flexoelectronic doping in semiconductors has opened a new direction in materials research, which can be applied to quantum, spintronics and electronics device applications.

**Author contributions**

PCL, AK and RGB have equal contribution to this work.

**Acknowledgement**

The fabrication of experimental devices was completed at the Center for Nanoscale Science and Engineering at UC Riverside. Electron microscopy imaging was performed at the Central Facility for Advanced Microscopy and Microanalysis at UC Riverside. SK acknowledges a research gift from Dr. Sandeep Kumar.

# Supplementary materials: Flexoelectronic doping of the degenerate silicon and correlated electron behavior

Paul C. Lou[1], Anand Katailiha[1], Ravindra G. Bhardwaj[1], Ward Beyermann[2], Dheeraj Mohata[3], and Sandeep Kumar[1,*]

[1] Department of Mechanical Engineering, University of California, Riverside, CA 92521, USA

[2] Department of Physics and Astronomy, University of California, Riverside, CA 92521, USA

[3] Global Communication Semiconductors LLC, Torrance, CA 90505

Supplementary Table S1. The flexoelectronic charge carrier transfer from Py to Si leading to change in charge carrier concentration in each layer and resulting resistivity behavior for each current estimated at 300 K.

| I | $\rho_{Py}$ (Ωm) | Ratio | $n_{Py}$ (cm$^{-3}$) | Reduction | $n_{Si}$ after flexoelectronic doping | Expected $\rho_{Si}$ (Ωm) | Calculated $\rho_{Si}$ (Ωm) | Error (%) |
|---|---|---|---|---|---|---|---|---|
| 10 μA | 6.90×10$^{-7}$ | 1.347 | 5.08×10$^{22}$ | 1.76×10$^{22}$ | 1.1×10$^{21}$ | 4.28×10$^{-7}$ | 1.33×10$^{-5}$ | 70.6 |
| 0.25 mA | 7.29×10$^{-7}$ | 1.424 | 4.81×10$^{22}$ | 2.04×10$^{22}$ | 1.28×10$^{21}$ | 3.71×10$^{-7}$ | 1.15×10$^{-5}$ | 81.6 |
| 0.5 mA | 7.37×10$^{-7}$ | 1.44 | 4.75×10$^{22}$ | 2.09×10$^{22}$ | 1.31×10$^{21}$ | 3.61×10$^{-7}$ | 8.65×10$^{-6}$ | 83.7 |
| 0.75 mA | 7.42×10$^{-7}$ | 1.45 | 4.72×10$^{22}$ | 2.13×10$^{22}$ | 1.33×10$^{21}$ | 3.56×10$^{-7}$ | 6.94×10$^{-6}$ | 85.0 |
| 1.0 mA | 7.47×10⁻⁷ | 1.46 | 4.69×10²² | 2.16×10²² | 1.35×10²¹ | 3.50×10⁻⁷ | 5.86×10⁻⁶ | 86.3 |
| 1.5 mA | 7.60×10⁻⁷ | 1.485 | 4.61×10²² | 2.24×10²² | 1.40×10²¹ | 3.38×10⁻⁷ | 4.42×10⁻⁶ | 89.5 |
| 2 mA | 7.70×10⁻⁷ | 1.504 | 4.55×10²² | 2.30×10²² | 1.44×10²¹ | 3.30×10⁻⁷ | 3.98×10⁻⁶ | 91.8 |

The temperature (T) dependent resistivity of the Py can be described by the following equation:

$$\rho_{Py}(T) = 3.289 \times 10^{-7} + 1.65 \times 10^{-10}T + 1.269 \times 10^{-12}T^2 \quad (S1)$$

The residual resistance of the composite sample is larger than the resistance of Py only thin film at 5 K, which is considered as the residual resistance of Py layer without any contribution form p-Si layer. Hence, the resistivity of Py at 300 K in the composite sample at 10 µA is $1.347\rho_{Py}(5\ K)$ based on the residual resistance at 5 K. The table S1 shows the resistivity of the composite sample at 300 K based on the multiplication factor estimated at 5 K. The column 3 represents the multiplication factor for each current dependent measurement. This factor is then used to calculate the new charge carrier concentration at 300 K shown in column 4 since the charge carrier concentration were measured at 300 K. We also estimate the reduction in charge carrier concentration (column 5) based on increase in resistivity. The number of charge carrier are then distributed to Si layer, which gives a new change carrier concentration of Si layer (column 6). Based on the column 6 data, we recalculate the resistivity of Si layer as shown in column 7. We, then subtract the Py resistance from the overall resistance to get the resistance and resistivity response of the p-Si layer, which is second to last column in the Supplementary Table S1. The difference in two resistivity based on two methods is given in the last column. This error can be attributed to the reduction in the mobility of the Si layer from flexoelectronic doping.

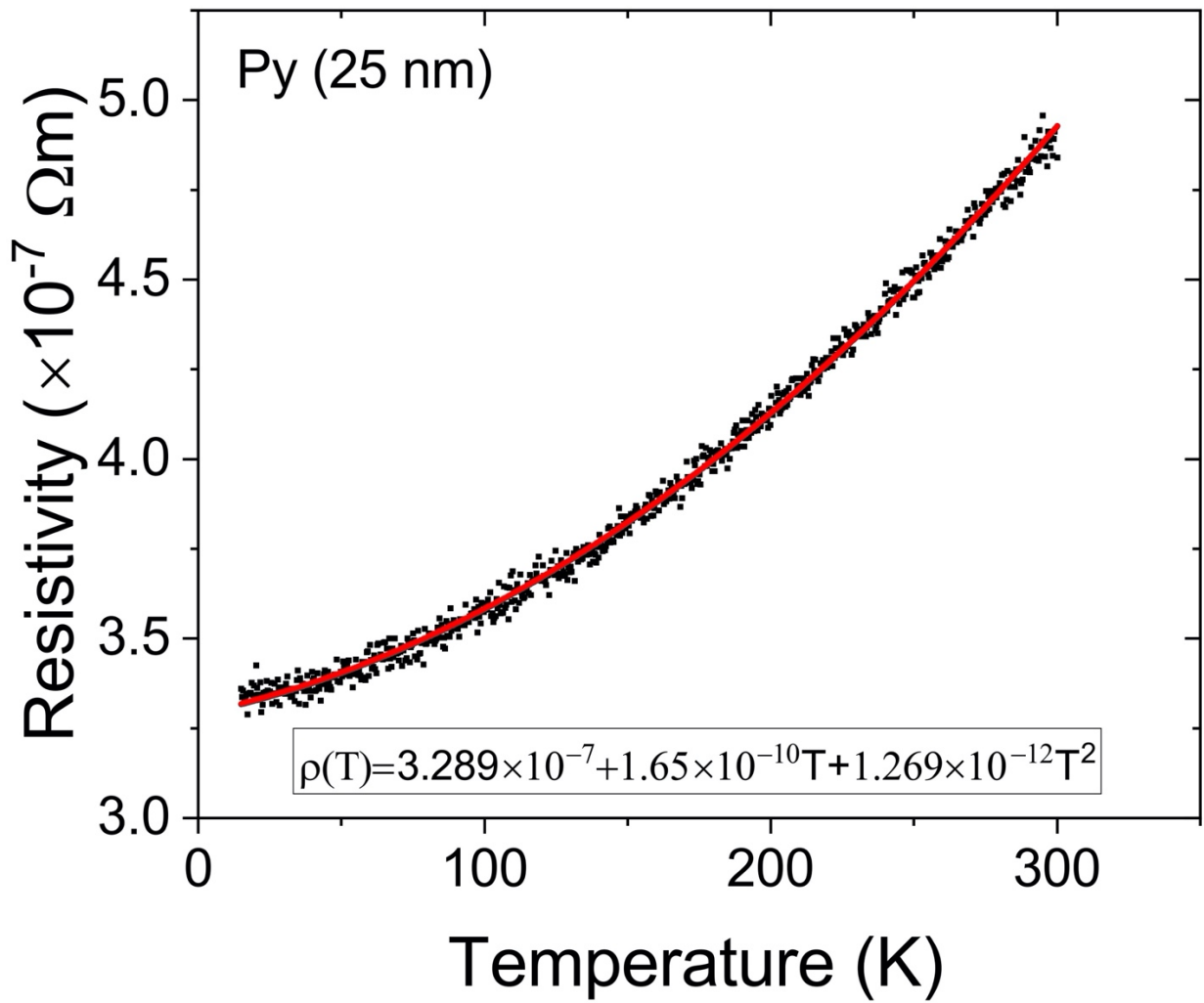

Supplementary Figure S1. The expected resistance behavior of Py layer as a function of temperature from 300 K to 15 K. The red line is the fit line.

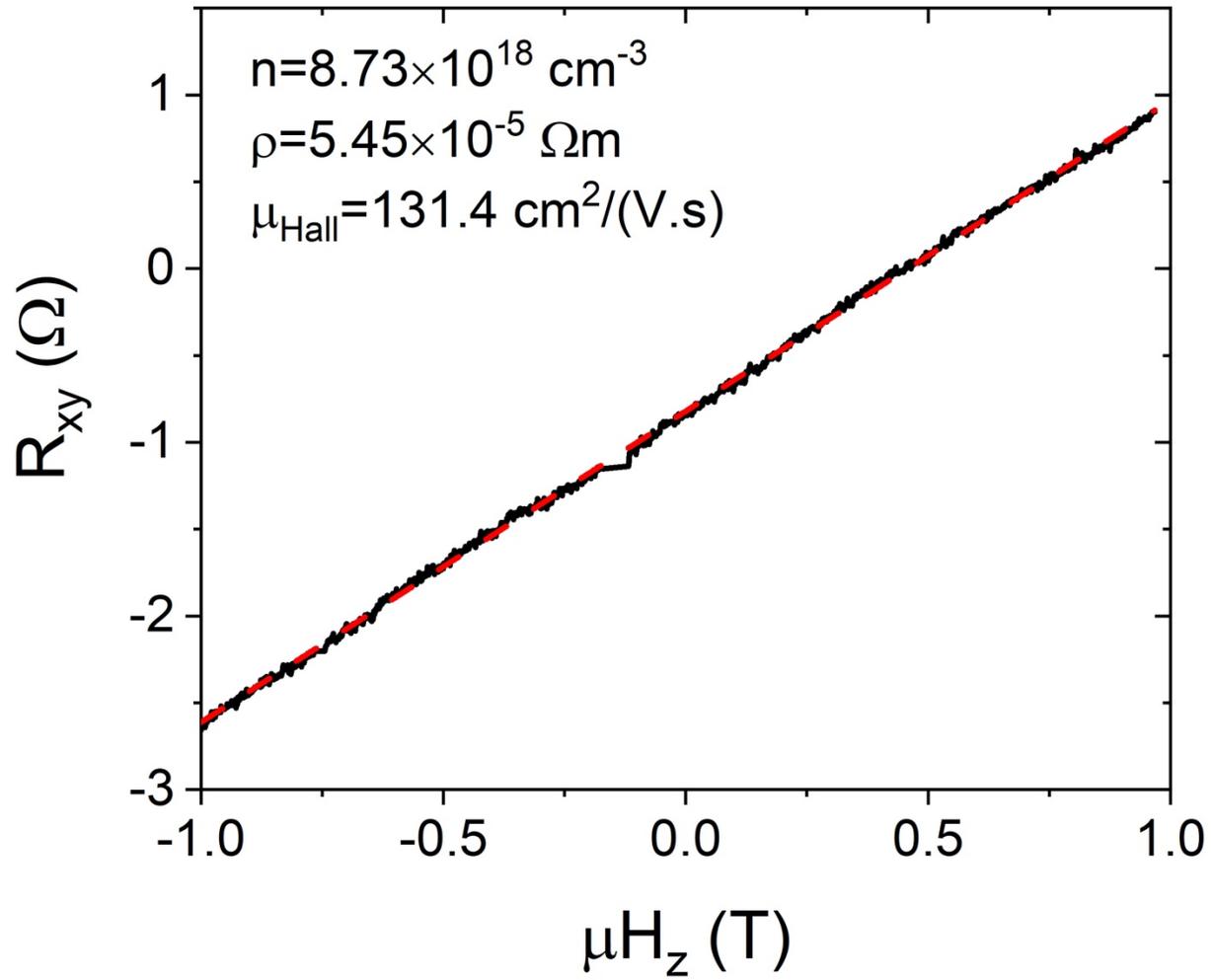

Supplementary Figure S2. The Hall response in 400 nm p-Si sample. The red line shows the line fit.